\documentclass[aps,prb,twocolumn,groupedaddress]{revtex4}
\usepackage{amssymb}

\usepackage{graphicx}
\usepackage{dcolumn}
\usepackage{bm}

\begin{document}

\title{Manifestation of the exchange enhancement of the valley splitting in
the quantum Hall effect regime}
\author{I. Shlimak}
\affiliation{Minerva Center and Jack and Pearl Resnick Institute of Advanced Technology,
Department of Physics, Bar-Ilan University, Ramat-Gan, Israel}
\author{K.-J. Friedland}
\affiliation{Paul-Drude Institut f\"{u}r Festk\"{o}rperelektronik, Hausvogteiplatz 5-7,
10117, Berlin, Germany}
\author{V. Ginodman}
\affiliation{Minerva Center and Jack and Pearl Resnick Institute of Advanced Technology,
Department of Physics, Bar-Ilan University, Ramat-Gan, Israel}
\author{S.~V. Kravchenko}
\affiliation{Physics Department, Northeastern University, Boston, Massachusetts 02115,
U.S.A.}
\date{\today}

\begin{abstract}
We report a new ``dip'' effect in the Hall resistance, $R_{xy}$, of a Si
metal-oxide-semiconductor field-effect transistor in the quantum Hall effect
regime. With increasing magnetic field, the Hall resistance moves from the
plateau at Landau filling factor $\nu =6$ directly to the plateau at $\nu =4$%
, skipping the plateau at $\nu =5$. However, when the filling factor
approaches $\nu =5$, the Hall resistance sharply ``dives'' to the value $%
1/5(h/e^{2})$ characteristic of the $\nu =5$ plateau, and then returns to $%
1/4(h/e^{2})$. This is interpreted as a manifestation of the oscillating
exchange enhancement of the valley splitting when the Fermi level is in the
middle between two adjacent valley-split Landau bands with the asymmetric
position of the extended states.
\end{abstract}

\maketitle

The quantum Hall effect (QHE) was discovered in Si metal-oxide-semiconductor
field-effect transistors (MOSFETs) more than 25 years ago.\cite{Klitzing}
This system is still under study, mainly because of the problem of the
metal-insulator transition in two-dimensional electron systems (2DES) (see,
for example, Ref. \cite{Kravchenko} and references therein). The main
interest of researches of the QHE is focused on much less disordered 2DES
based on Si/SiGe and GaAs/AlGaAs structures with high electron mobility. We
are aware of only a few publications where the Hall resistivity $\rho _{xy}$
was measured in Si-MOSFET in a narrow interval of magnetic fields $B$ and
gate voltages $V_{\mathrm{g}}$.\cite{Pudalov, Khrapai, Cheremisin}

In the present work, we report the results of transport measurements in two
Si-MOSFET samples in the QHE regime in a wide interval of $B$ (up to 14~T)
and electron densities $n$ (up to $1.5\cdot10^{16}~\mathrm{m}^{-2})$,
controlled by $V_\mathrm{g}$.

% Both samples were prepared from the same wafer and had similar parameters.
% *** i thik they are not from the same wafer... and parameters are not that close.  So let's just omit it.
Measurements of $n$ and electron mobility $\mu $ at $T=0.3$~K in sample \#1
yield a linear dependence $n(V_{\mathrm{g}})=1.41\cdot 10^{15}$ $(V_{\mathrm{%
g}}-0.4~\mathrm{V})~\mathrm{m}^{-2}$ within the interval of $V_{\mathrm{g}}$
from 5.5 to 11~V with $\mu \approx 1.0\div 1.5~\mathrm{m^{2}/V\cdot s}$.
Sample \#2 was measured in the interval $V_{\mathrm{g}}$ between 8 and 11~V,
and $n(V_{\mathrm{g}})$ was approximated by a different linear dependence $%
n(V_{\mathrm{g}})=1.25\cdot 10^{15}$ $(V_{\mathrm{g}}+0.6~\mathrm{V})~%
\mathrm{m}^{-2}$, with very similar values for $n$ in the measured interval
of $V_{\mathrm{g}}$. The mobility in sample \#2 was also within the above
interval, changing from $\mu =1.46~\mathrm{m^{2}/V\cdot s}$ at $V_{\mathrm{g}%
}=8$~V to $1.27~\mathrm{m^{2}/V\cdot s}$ at $V_{\mathrm{g}}=11$~V. The
sample resistance was measured using a standard lock-in technique with the
measuring current 20 nA at a frequency of 10.6 Hz.

\begin{figure}[b]
\includegraphics[width=8cm]{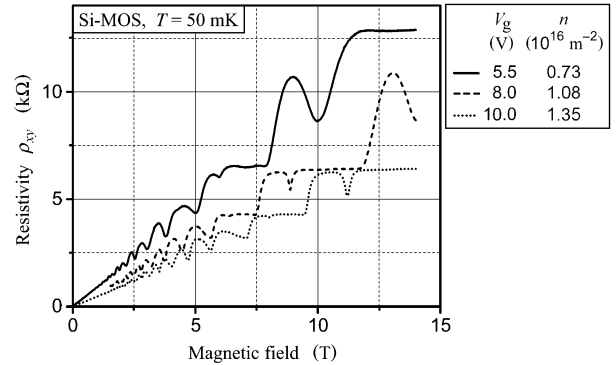}
\caption{Dependence of the Hall resistance $\protect\rho_{xy}$ on the
perpendicular magnetic field $B$ for different gate voltages $V_\mathrm{g}$.
The electron density is as indicated.}
\label{}
\end{figure}

Figure 1 shows $\rho _{xy}$ of sample \#1 as a function of perpendicular
magnetic field $B$ at different $V_{\mathrm{g}}$. One can see two features.
The first one is the ``overshoot'' effect which is observed at almost every
plateau, being especially large at filling factor $\nu =3$ (Fig.~2). In
incremental magnetic fields, when $\nu $ approaches the integer 3, $\rho
_{xy}$ overshoots the normal plateau value $1/3(h/e^{2})=8.6$~kOhm. However,
as $B$ increases further, $\rho _{xy}$ drops to its normal value. The
overshoot effect has been previously observed in GaAs/AlGaAs and Si/SiGe
heterostructures (see, for example, Ref. \cite{ShlimakPRB} and references
therein).

\begin{figure}[tb]
\includegraphics[width=8cm]{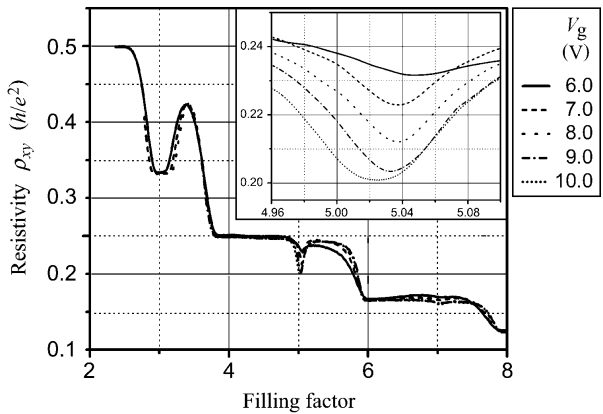}
\caption{Dependence of the dimensionless resistivity $\protect\rho%
_{xy}/(h/e^2)$ on the filling factor $\protect\nu=nh/eB$ for different gate
voltages $V_\mathrm{g}$. The insert shows the enhanced view of the ``dip''
effect.}
\label{}
\end{figure}

The second feature is new. It consists of a ``dip'' of $\rho _{xy}$ from the
plateau at $\nu =4$ (6.45~kOhm) to the plateau at $\nu =5$ (5.16~kOhm) at
magnetic fields when the filling factor approaches $\nu =5$. This effect can
be seen more clearly in Fig.~2, where the dimensionless Hall resistivity (in
units of $h/e^{2}$) is plotted as a function of the filling factor $\nu
=nh/eB$.

Figure 3 shows that the longitudinal resistivity $\rho_{xx}$ also exhibits a
``dip'' at $\nu=5$ and a less pronounced ``dip'' at $\nu=7$.

In the present work, the ``dip'' effect was invariably observed in all
experiments, in both samples, for different voltage probes and reversed
directions of the current and magnetic field. Moreover, Figs. 1 and 2 show
the development of the ''dip'' with variation of the gate voltage for the
same sample and probes. These facts give us confidence that the observed
''dip'' is not connected with heterogeneity of the 2DES and possible
admixture of $R_{xx}$ into $R_{xy}$. Let us discuss the origin of the
''dip'' effect.

For Si-based 2DES, like Si/SiGe and Si-MOSFETs, the energy spectrum in a
magnetic field is 
\[
E_{n}=\hbar \omega _{\mathrm{c}}(N+\frac{1}{2})\pm \frac{\Delta E_{\mathrm{s}%
}}{2}\pm \frac{\Delta E_{\mathrm{v}}}{2}
\]%
where $N=0,1,\ldots $ is the Landau level (LL) number, $\omega _{\mathrm{c}%
}=eB_{\bot }/mc$ is the cyclotron frequency, $\Delta E_{\mathrm{s}}=g^{\ast
}\mu _{\mathrm{B}}B$, is the Zeeman splitting, $g^{\ast }$ is the effective
Land\'{e} factor, $\mu _{\mathrm{B}}=e\hbar /2m_{e}$ is the Bohr magneton, $%
B=(B_{\bot }^{2}+B_{\Vert }^{2})^{1/2}$ is the total magnetic field. $\Delta
E_{\mathrm{v}}~[K]=\Delta _{\mathrm{v}}^{0}+0.6B_{\bot }~[T]$ is the valley
splitting energy, $\Delta _{\mathrm{v}}^{0}$ is assumed \cite%
{Pudalov,Cheremisin} to be 2.4~K or 0.9~K. In accordance with this scheme,
odd filling factor corresponds to the Fermi level position $\varepsilon _{F}$
midway between two adjacent valley-split LL (see insert in Fig.~4). If one
takes into account the disorder in real samples, each LL is broadened into a
Landau band (LB), with the width determined by the scale of the disorder
energy $W$.

In Fig. 4, the dependences of $\rho _{xy}(\nu )$ for $n$-Si/SiGe \cite%
{ShlimakPRB, ShlimakEPL} and for $n$-Si-MOSFET with almost the same electron
concentration $n$ are shown together for comparison. In Si/SiGe, all
plateaus are clearly observed, including plateaus at odd filling factor. In
Si-MOSFET, plateaus at odd filling factors with $\nu >3$ are not observed.
This can be explained by the increase of disorder in Si-MOSFET, because the
interface between Si and SiO$_{2}$ is much less perfect than the interface
between Si and SiGe. Increase of disorder results in the broadening of LB in
Si-MOSFET. If the width of LB is of order of the valley-splitting energy $%
\Delta E_{\mathrm{v}}$, the density-of-states $N(\varepsilon )$ does not
have a deep minimum between valley-split adjacent Landau bands, the Fermi
level does not linger between them, and the corresponding plateau is missed.

\begin{figure}[tb]
\includegraphics[width=8cm]{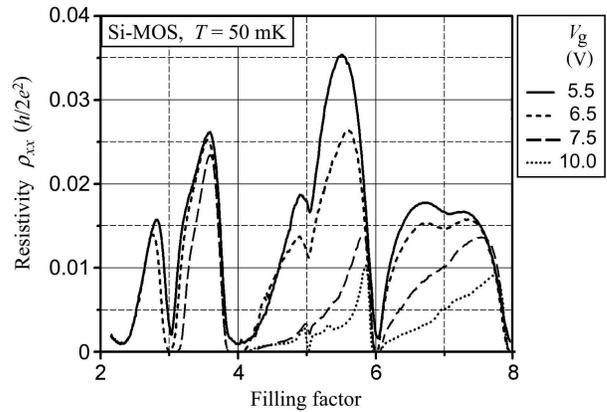}
\caption{Longitudinal resistivity in dimensionless units $\protect\rho%
_{xx}/(h/2e^{2})$ as a function of filling factor $\protect\nu =nh/eB$ for
different $V_{\mathrm{g}}$.}
\end{figure}

\begin{figure}[b]
\includegraphics[width=7cm]{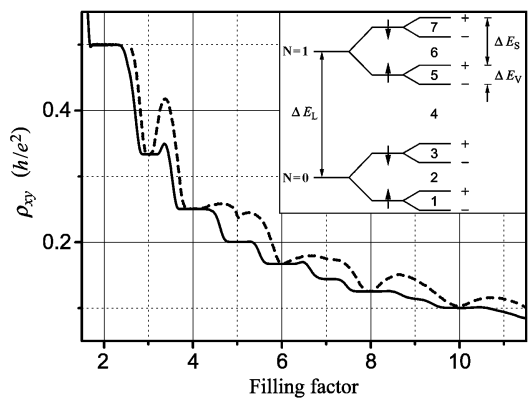}
\caption{Comparison of QHE in Si/SiGe ($n=8.94\cdot 10^{15}~\mathrm{m}^{-2}$%
, solid line) and in Si-MOSFET with similar electron concentration ($%
n=8.80\cdot 10^{15}~\mathrm{m}^{-2}$, dashed line).}
\label{}
\end{figure}

However, when $\nu $ approaches the integer $\nu =5$, the value of $\rho
_{xy}$ starts to fall toward the missed plateau and finally reaches this
plateau with increasing $n$ (see insert in Fig.~2). This effect can be
explained by the temporary enhancement of the valley splitting. It was shown %
\cite{Ohkawa,Rauh,Ando} that the occupied Landau levels undergo a
self-energy shift roughly proportional to their occupation and inversely
proportional to the screening of the system. Therefore, the enhancement
oscillates as a function of the electron occupation of LB and has the
maximum value when the filling factor approaches an integer and the Fermi
level lies midway between the adjacent valley-split LB (see Fig.~5). We
believe that this exchange enhancement of the valley splitting is
responsible for the ``dip'' effect at $\nu =5$.

The model of spin-split exchange enhancement was used to explain the
enhanced $g$-factor in GaAs/AlGaAs\cite{Richter} and the bistable switching
between quantum Hall conduction and dissipative conduction near $\nu =1$ in
a quantum Hall system in GaInAs quantum well.\cite{Nachtwei} Exchange
enhancement was also used to explan the overshoot effect at $\nu =3$ in
Si/SiGe heterostructure.\cite{Weitz} However, overshoot is observed at the
low magnetic-field edge of the $\rho _{xy}$ plateau when $\nu $ is far from
an integer. Furthermore, after overshoot, $\rho _{xy}$ remains at its
``normal'' plateau value, while the enhanced splitting due to exchange
interaction oscillates and has maximum at the integer $\nu $. Therefore, the
manifestation of the exchange enhanced splitting is expected in the close
vicinity of integer $\nu $ in the form of a ``dip'', which is observed in
our experiment for the first time (Fig.~2).

We would like to mention that non-monotonic behavior of the Hall resistance
with several reentrances of the plateau values 1/4$(h/e^2)$ and 1/3$(h/e^2)$
was observed in a modulation-doped GaAs quantum well at very low tempertures
(15~mK) in Ref.\cite{Eisenstein} and explained by the existence of
collective insulating states in the N = 1 Landau level. Most likely, physics
behind this phenomenon and our ``dip'' is entirely different, although the
two effects look similar.

One can see also from Fig.~2 that in Si-MOSFET, the integer values of $\nu $
do not correspond to the middle point of the plateaus, in contrast with more
perfect Si/SiGe (Fig.~4). This can be considered as evidence of asymmetry of
LB in Si-MOSFET, when delocalized states are displaced from the center of LB
(Fig.~5). As a result of this asymmetry and considerable overlap of the
adjacent valley-splt LB, the Fermi level $\varepsilon _{F}$ is situated in
the interval of localized states corresponding to the plateau with $\rho
_{xy}=1/4$ (in units of $h/e^{2})$ even at $\nu \gtrsim 5$, Fig.~5(a).
However, as $\varepsilon _{F}$ approaches exactly $\nu =5$, the valley
splitting increases leading to LB separation, Fig.~5(b). The localized
states, which correspond to the plateau 1/5, show up, and $\rho _{xy}$ dives
to 1/5. When $\nu $ is further decreased, the exchange interaction-induced
enhancement of the valley splitting disappears and $\varepsilon _{F}$ finds
itself again in the interval of localized states which corresponds to the
plateau $\rho _{xy}=1/4$, Fig. 5(c).

Asymmetry in the position of the delocalized states could be a consequence
of the asymmetry of large potential fluctuations caused by the fact that an
excess of the local electron concentration above the average value $\langle
n\rangle $ is, in principle, unlimited, while the local deficit of electron
density is limited by the value of $\langle n\rangle $ itself. To satisfy
neutrality, the area occupied by the negatively charged fluctuations is less
than the area of the positively charged fluctuations. Correspondingly, the
integral number of localized states above the percolation level is less than
the integral number of states below this level which explain the asymmetry
of the position of delocalized states in disorder-broadened LB. Computer
simulation also shows that the increase of disorder leads to the
displacement of delocalized states from the central part of LB.\cite{Aldea}
In more perfect systems, potential fluctuations are small, and asymmetry is
negligible, which explain why in Si/SiGe, the integer values of $\nu $
correspond approximately to the middle point of each plateau.

The question arises why the ``dip'' effect is clearly observed in Si-MOSFET
at high electron densities in relatively strong magnetic fields but barely
observed at low electron densities (Fig.~1); why it does not exist in
Si/SiGe? We believe that the necessary condition for observation of the
``dip'' effect is the equality of the splitting energy $\Delta E$ and the
width of the adjacent LB. The last parameter could be estimated roughly as
the average energy of disorder $W$. In more perfect systems, $W\ll \Delta E$
and the Fermi level is fixed between the narrow adjacent LB even without
enhanced splitting. As a result, the odd plateaus are clearly observed in
Si/SiGe. In the opposite case, $W\gg \Delta E$ (Si-MOSFET with low electron
density), the adjacent strongly broadened valley-split LB remain unresolved
even in the case of enhanced splitting. The valley-splitting is small in
weak magnetic fields, which explains why the ``dip'' effect is barely
observed at $\nu =7$.

\begin{figure}[tb]
\includegraphics[width=8cm]{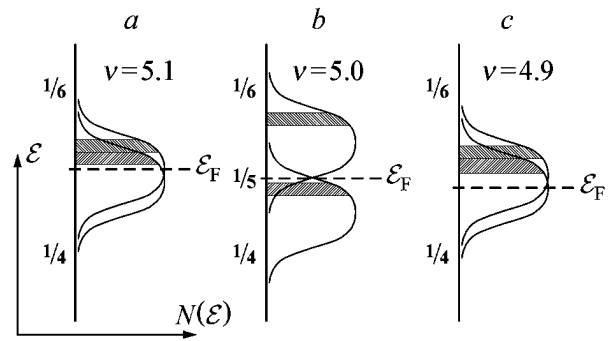} \label{}
\caption{Schematic sketch of the splitting enhancement between adjacent
valley-split LB when the filling factor passes over $\protect\nu =5$.
Delocalized states, shown as shaded bands, are displaced from the center of
LB. Fractional numbers correspond to the plateau resistance in units $h/e^2$%
. Magnetic field increases from \emph{a} to \emph{c}.}
\end{figure}

What follows from these considerations is that in our case, $W\approx \Delta
E_{\mathrm{v}}$. One can roughly estimate the average energy of disorder as $%
W\approx \hbar /\tau $, where $\tau $ is the time between elastic collisions
which determines the mobility $\mu =e\tau /m^{\ast }$. In our sample, $\mu
=1.0\div 1.5~\mathrm{m^{2}/V\cdot s}$. Using $m^{\ast }=0.2m_{0}$ for
strained Si layers,\cite{Ando} we obtain $W\approx 7\div 10$~K. Figure~1
shows that the ``dip'' effect is clearly observed at $B\approx 10\div 12$~T.
In these fields, $\Delta E_{\mathrm{v}}\approx 8\div 10$~K, which is indeed
equal to the above estimate of $W$ and confirms our model.

It was also predicted \cite{Rauh, Ando} that the exchange enhancement drops
drastically around a certain temperature due to a two-fold positive feedback
mechanism: with increasing temperature the difference in occupation numbers
of the two valleys decreases and simultaneously the screening increases.

This effect was observed in our experiment. Figure 6 shows the temperature
dependence of the ``dip'' at $\nu =5$ measured at $V_{\mathrm{g}}=11$~V and $%
B=12.5$~T plotted in the Arrhenius scale. The maximum amplitude of the
``dip'' is the difference between two plateaus at 1/4 and 1/5 in units of $%
h/e^2$ and is obtained at low temperatures. With increasing $T$, the
amplitude of ``dip'' decreases first weakly but at $T>1$~K the amplitude
drops drastically with the energy of activation $T_0=11\pm 3$~K, which is
indeed approximately equal to the valley splitting $\Delta E_{\mathrm{v}}$
at $B=12.5$~T.

\begin{figure}[tb]
\includegraphics[width=8cm]{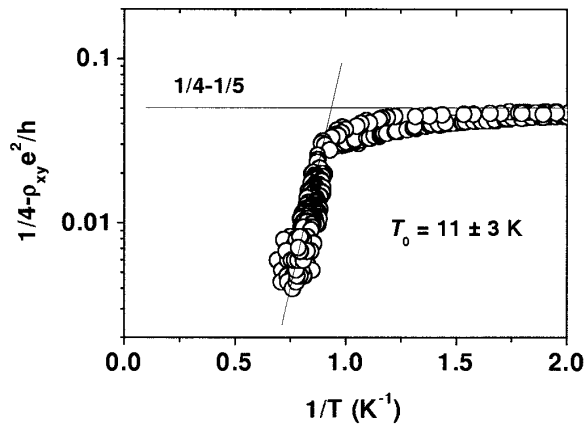}
\caption{Temperature dependence of the ``dip'' amplitude at $\protect\nu=5$.}
\label{}
\end{figure}

In summary, we have observed a new ``dip'' effect in the Hall resistivity of
Si-MOSFET measured in the quantum Hall effect regime. This effect can be
considered as a manifestation of the oscillating enhancement of the valley
splitting due to the exchange interactions. Observation of the ``dip''
effect is preferable if the width of the adjacent LB is approximately equal
to the initial valley-split energy.

We thank V.~T. Dolgopolov, Y.~Levinson, and A.~Finkel'stein for fruitful
discussions and the Erick and Sheila Samson Chair of Semiconductor
Technology for financial support. \vspace*{\fill}

%\bibliography{apssamp} 

\end{document}